\documentstyle[aps,epsf,twocolumn]{revtex} 
\begin{document} \def\inseps#1#2{\def\epsfsize##1##2{#2##1}
\centerline{\epsfbox{#1}}}
\twocolumn[\hsize\textwidth\columnwidth\hsize\csname@twocolumnfalse\endcsname

\title{The r\^ole of the Berry Phase in Dynamical Jahn-Teller Systems}
\author{Nicola Manini} \address{European Synchrotron Radiation Facility,
B.P. 220, F-38043 Grenoble C\'edex, France}
\author{Paolo De Los Rios} \address{Institut de Physique Th\'eorique,
Universit\'e de Fribourg, 1700-CH Fribourg, Switzerland}

\date{\today} \maketitle

\begin{abstract}
The presence/absence of a Berry phase depends on the topology of the
manifold of dynamical Jahn-Teller potential minima.  We describe in detail
the relation between these topological properties and the way the lowest
two adiabatic potential surfaces get locally degenerate.  We illustrate our
arguments through spherical generalizations of the linear $T \otimes h$ and
$H \otimes h$ cases, relevant for the physics of fullerene ions.  Our
analysis allows us to classify all the spherical Jahn-Teller systems with
respect to the Berry phase.  Its absence can, but does not necessarily,
lead to a nondegenerate ground state.
\end{abstract}

\pacs{PACS numbers: 31.90.+s,71.20.Tx,71.38.+i}

\vspace{1cm} \vfill ] \narrowtext

\section{Introduction}

The traditional field of degenerate electron-lattice interactions
(Jahn-Teller effect) in molecules and impurity centers in
solids\cite{Englman,Bersuker} has attracted new interest in recent years,
excited by the realization of new systems which call for a revision of a
number of commonly accepted beliefs.  A whole range of icosahedral
molecular systems including C$_{60}$ ions and some higher fullerenes,
thanks to the rich structure of the symmetry group, are characterized by up
to fivefold-degenerate representations of the electronic and vibrational
states of the isolated molecule/ion.  New Jahn-Teller (JT) systems have
therefore been considered theoretically,\cite{Bersuker,Ihm,AMT} with
intriguing features.\cite{AMT,Mead,Wilczek,Delos96}  A particularly
surprising property has been demonstrated recently: the possibility of a
symmetry switch of the molecular ground state\cite{Delos96,Moate96},
connected to the {\em absence} of Berry phase in the coupled dynamics of
electrons and vibrations.

As it is well known, the molecular symmetry, reduced in the static JT
effect with the splitting of the electronic-state degeneracy, is restored
when the coherent tunneling between equivalent distortions is considered,
in the dynamical Jahn-Teller (DJT) effect.  In this context it was commonly
accepted an empirical ``symmetry conservation rule'', sometimes referred to
as ``Ham's theorem'', stating that the symmetry of the vibronic DJT ground
state, at all coupling strengths, remains the same as that of the
electronic multiplet prior to coupling:\cite{Bersuker} all linear JT
systems known till a few years ago, for single-electron occupancy,
systematically satisfy this empiric rule.  It was understood recently that
this phenomenon, not automatically implied by the DJT physics, is in
reality a fingerprint of a Berry phase\cite{Berry} in the entangled
electronic-phononic dynamics.\cite{AMT,Delos96,ob96}  Consequently, this
geometrical phase seemed a universal feature of the DJT systems.

It came unexpected the discovery of the first dynamical JT system
{\em without} Berry phase, showing a {\em nondegenerate ground state} in
the strong-coupling limit.\cite{Delos96,Moate96}  This is the model that in
spherical symmetry is indicated as ${\cal D}^{(2)}\otimes d^{(2)}$, where
electrons of angular momentum $L=2$ interact with vibrations also belonging
to an $l=2$ representation.  This system is relevant to the physics of
fullerene ions C$_{60}^+$, where the 5-fold degenerate electronic state has
$H_u$ icosahedral label and the quadrupolar distortions correspond to some
of the $h_g$ modes.\cite{Delos96}  It has been shown both analytically and
numerically that, for increasing coupling, a nondegenerate state in the
vibronic spectrum moves down, to cross the 5-fold ground state at some
finite value of the coupling parameter, thus becoming the ground state at
strong coupling~\cite{Delos96,Moate96}.

In this work, we review the mechanism of the Berry phase in degenerate
electron-vibration coupled systems, and uncover the detailed reason of the
absence of this geometric phase in the ${\cal D}^{(2)}\otimes d^{(2)}$
system, hence drawing some natural generalizations.  We structure the paper
as follows. In Sect.\ \ref{models:sect}, we introduce the basic JT
Hamiltonian, along with the fundamental transformations to recast it in a
semi-classical form.  Section \ref{bp:sect} reviews the mechanism
responsible for the presence of the Berry phase in most DJT systems, and
its consequences for the low-energy part of their spectrum. In Sect.\
\ref{nobp:sect}, the mechanism allowing to get rid of the Berry phase in
some systems is unveiled, and a class of models sharing this property is
proposed.  Section \ref{numerical:sect} reports some numerical results
confirming and complementing the predictions of Sect.\ \ref{nobp:sect}.  In
Sect.\ \ref{discussion:sect}, we discuss possible generalizations of our
model to higher-order couplings and discrete symmetry groups.  Finally,
conclusions are drawn in Sect.\ \ref{conclusion:sect}.

\section{The models}
\label{models:sect}
According to the general theory of the JT effect,\cite{Bersuker} an
$N$-fold degenerate electronic level corresponding to a representation
$\Gamma$ of the molecular symmetry point group interacts with all the
vibrational modes corresponding to representations $\{\Lambda\}$ contained
in the symmetric part of the direct product of $\Gamma$ with itself.  For
simplicity, in this work we restrict to only one degenerate mode.  The
Hamiltonian for this ``$\Gamma \otimes \Lambda$'' case reads:
\begin{equation}
H = \hbar \omega \sum_{i=1}^{|\Lambda|}
b^{\dagger}_i b_i + H_{\rm e-v}
\label{2nd quant ham}
\end{equation}
where $b^{\dagger}_i$ / $b_i$ are the creation/destruction operators for
the harmonic vibrational mode component $i$, and $H_{\rm e-v}$ is the
interaction Hamiltonian, which, to linear order in the boson operators, writes
\begin{equation}
H_{\rm e-v} = \frac 12
g \hbar \omega				
\sum_{i=1}^{|\Lambda|} \sum_{j,k=1}^{|\Gamma|}
\left(b_i c^\dagger_j c_k
\langle\Lambda i | \Gamma j \Gamma k\rangle + h.c. \right)\;,
\label{interaction Hamiltonian}
\end{equation}
$c_j$ is the fermion operator for orbital $j$, $\langle \Lambda i | \Gamma
j \Gamma k\rangle$ are the Clebsch-Gordan coefficients of the symmetry
group of the problem, and $g$ is the dimensionless coupling parameter.  For
the purpose of illustrating our analysis, it is particularly convenient to
stick to the case in which the symmetry group is that of three-dimensional
rotations, SO(3).  In the following, therefore, we focus on the coupling of
an electronic state of angular momentum $L$ (whose representation
$\Gamma$ is indicated as ${\cal D}^{(L)}$) to a degenerate vibration of
angular momentum $l$ (of representation $\Lambda\equiv {d}^{(l)}$).
For this case, the Hamiltonian reads
\begin{eqnarray}
H & = &
\hbar \omega \sum_{m=-l}^{l}\left\{
b^{\dagger}_m b_m + \frac g2
\sum_{k,k'=-L}^{L} \left(-1\right)^{k'}
\nonumber \right.\\
& &
\left.
\left[b^\dagger_m + \left(-1\right)^m b_{-m}\right] c^\dagger_{k} c_{k'}
\langle l m | L k, L -k' \rangle\  \right\} \;,
\label{ham sferica}
\end{eqnarray}
where some of the symmetries of the $SO(3)$ Clebsch-Gordan coefficients are
implied.

The Hamiltonian (\ref{ham sferica}) is suitable for perturbation
calculations (small $g$ values) and as a starting point for numerical
diagonalization methods, such as the Lanczos technique.  Yet, the Berry
phase is a semiclassical concept, useful only in the medium/large $g$
regimes. Here it is convenient to switch to a {\it real} representation of
the vibrational degrees of freedom.

We apply two unitary transformations, both on the electronic operators, and
the vibrational ones.  We define a new set of electronic operators,
$\tilde{c}_m$ (and consequently their hermitian conjugate
$\tilde{c}_m^\dagger$), $m=-L,...,L$ via the transformation
\begin{eqnarray}
\label{ele trans}
\tilde{c}_0&=&c_0 \\
\pmatrix{ \tilde{c}_m \cr \tilde{c}_{-m} }
&=& \frac{
\exp \frac{i \pi \left[(-)^m -1\right]}{4}
}{\sqrt{2}}
\left(\begin{array}{cc} i & -i \\ 1 & 1 \end{array} \right)
\pmatrix{ c_m \cr c_{-m} }, \  m>0~. \nonumber
\end{eqnarray}
The second transformation expresses the $2 l +1$ boson operators $B_m\equiv
b^\dagger_m + \left(-1\right)^m b_{-m}$ in terms of the Hermitean
``coordinate'' operators $q_m$ as follows:
\begin{eqnarray}
\label{coordinate}
B_0 & = & \sqrt{2} q_0
\\
\pmatrix{B_m \cr B_{-m} }
&=&
\left(\begin{array}{cc} (-)^m & -i (-)^m \\ i & 1 \end{array} \right)
\pmatrix{ q_m \cr q_{-m} }, \  m>0~. \nonumber
\end{eqnarray}
The remaining components $b^\dagger_m - \left(-1\right)^m b_{-m}$ can be
expressed correspondingly in terms of the momentum operators $p_m$
conjugate to $q_m$.  We stress that these transformations are by no means
unique.  Any further orthogonal transformation of the real coordinates,
leads to equivalent results.

Eventually, the Hamiltonian operator (\ref{ham sferica}) transforms into the
form
\begin{equation}
H = \frac{1}{2} \hbar \omega_{l}
\sum_{M=-l}^{l} (p_{m}^2 + q_{m}^2) + H_{\rm e-v}
\label{real Hamiltonian}
\end{equation}
with
\begin{equation}
H_{\rm e-v} = \frac {1}{\sqrt{2}}
g \hbar \omega \sum_{m=-l}^l q_m
\sum_{j,k=-L}^{L}
\tilde{c}^\dagger_{j}  V^{(m)}_{j,k} \tilde{c}_{k} \; .
\label{real interaction Hamiltonian}
\end{equation}
It is straightforward to compute the $(2 L+1)\times(2 L+1)$ coupling
matrices $ V^{(m)}$, for any value of $L$ and l.  For brevity, we introduce
the symbols $\gamma^{+}_{j,k}\equiv\langle l\;j+k | L\;j, L\;k \rangle$,
$\gamma^{-}_{j,k} \equiv 0$.  In this notation, the matrices for the $L=2$
case are:
\begin{eqnarray}
\label{V0}
V^{(0)}&=&
\left(\begin{array}{ccccc}
\gamma^{+}_{-2,2} &0 &0 &0 & 0 \\
0 &-\gamma^{+}_{-1,1}&0 &0 & 0 \\
0 &0 & \gamma^{+}_{0,0} &0 & 0 \\
0 &0 &0 &-\gamma^{+}_{1,-1}& 0 \\
0 &0 &0 &0 &\gamma^{+}_{2,-2}
\end{array} \right) \\
\label{V1}
V^{(\pm 1)}&=&
\left(\begin{array}{ccccc}
0 &-\frac{\gamma^{\mp}_{-1,2}}{\sqrt{2}}&0 &
			\frac{\gamma^{\pm}_{-1,2}}{\sqrt{2}} & 0 \\
-\frac{\gamma^{\mp}_{2,-1}}{\sqrt{2}}&0 &-\gamma^{\mp}_{0,1} & 0 &
					-\frac{\gamma^{\pm}_{2,-1}}{\sqrt{2}}\\
0 &-\gamma^{\mp}_{1,0}&0 &-\gamma^{\pm}_{1,0} & 0 \\
\frac{\gamma^{\mp}_{2,-1}}{\sqrt{2}}&0 &-\gamma^{\pm}_{0,1} & 0 &
				-\frac{\gamma^{\mp}_{2,-1}}{\sqrt{2}}\\
0 &-\frac{\gamma^{\pm}_{-1,2}}{\sqrt{2}}&0 &
			-\frac{\gamma^{\mp}_{-1,2}}{\sqrt{2}} & 0
\end{array} \right) \\
\label{V2}
V^{(\pm 2)}&=&
\left(\begin{array}{ccccc}
0 &0 &-\gamma^{\pm}_{0,2} & 0 & 0 \\
0 &-\frac{\gamma^{\pm}_{1,1}}{\sqrt{2}}&0 &
		\frac{\gamma^{\mp}_{1,1}}{\sqrt{2}} & 0 \\
\gamma^{\pm}_{0,2}&0 &0 & 0 &-\gamma^{\mp}_{2,0}\\
0 &\frac{\gamma^{\mp}_{1,1}}{\sqrt{2}}&0 &
		\frac{\gamma^{\pm}_{1,1}}{\sqrt{2}} & 0 \\
0 &0 &-\gamma^{\mp}_{0,2} & 0 & 0
\end{array} \right) \\
\label{V3}
V^{(\pm 3)}&=&
\left(\begin{array}{ccccc}
0 &-\frac{\gamma^{\mp}_{1,2}}{\sqrt{2}}&0 &
			-\frac{\gamma^{\pm}_{1,2}}{\sqrt{2}} & 0 \\
-\frac{\gamma^{\mp}_{2,1}}{\sqrt{2}}&0 &0 & 0 &
			-\frac{\gamma^{\pm}_{2,1}}{\sqrt{2}}\\
0 & 0 &0 &0 & 0 \\
-\frac{\gamma^{\pm}_{2,1}}{\sqrt{2}}&0 &0 & 0 &
				\frac{\gamma^{\mp}_{2,1}}{\sqrt{2}}\\
0 &-\frac{\gamma^{\pm}_{1,2}}{\sqrt{2}}&0 &
			\frac{\gamma^{\mp}_{1,2}}{\sqrt{2}} & 0
\end{array} \right) \\
\label{V4}
V^{(\pm 4)}&=&
\left(\begin{array}{ccccc}
\frac{\gamma^{\pm}_{2,2}}{\sqrt{2}}&0 &0 & 0 &
		-\frac{\gamma^{\mp}_{2,2}}{\sqrt{2}}\\
0 & 0 &0 &0 & 0 \\
0 & 0 &0 &0 & 0 \\
0 & 0 &0 &0 & 0 \\
-\frac{\gamma^{\mp}_{2,2}}{\sqrt{2}}&0 &0 & 0 &
		-\frac{\gamma^{\pm}_{2,2}}{\sqrt{2}}\\

\end{array} \right)
\end{eqnarray}
Note that the forms of $V^{(0)}$, $V^{(\pm 1)}$, and $V^{(\pm 2)}$ apply to
the coupling to both $l=2$ and $l=4$ vibrons (of course, the numerical
values of the coefficients are different), while the $V^{(\pm 3)}$ and
$V^{(\pm 4)}$ matrices are relevant only for $l=4$.

For larger $L$'s, the structure of the central $5\times 5$ block
(corresponding to indexes $j=-2$ to 2) of the corresponding coupling
matrices is conserved, additional matrix elements being added externally in
a simple way.  For example, for any $L$ and $l$, the $V^{(0)}$ matrix is
diagonal with matrix elements $V^{(0)}_{k, k}=(-1)^k \gamma^+_{k,-k}$.
Similarly, if $l=2\cdot L$ then the only nonzero elements of the $V^{(\pm
l)}$ matrices are $V^{(l)}_{\pm l,\pm l}=\pm
\frac{\gamma^{\pm}_{l,l}}{\sqrt{2}}$ and $ V^{(-l)}_{\pm l,\mp l}=-
\frac{\gamma^{\pm}_{l,l}}{\sqrt{2}}$.

As anticipated, the form (\ref{real Hamiltonian},\ref{real interaction
Hamiltonian}) of the Hamiltonian is suitable to a semiclassical treatment,
in the spirit of the Born-Oppenheimer (BO) approximation, i.e.\ a
factorization of the electronic ``fast'' dynamics from the ``slow''
distortions $q_m$, which are quantized as a second step.  In this scheme,
we briefly review the Berry phase mechanism, and its consequences on the
factorized dynamics.

\section{The Berry Phase}
\label{bp:sect}

The traditional BO scheme assumes that the electrons, moving much faster
than the ions, follow adiabatically the ionic (vibrational) motion with the
only effect of generating a potential energy for the vibrational motion.
This approximation relies on separations between consecutive
electronic levels that are
much larger than the typical vibrational energies $\hbar \omega$.  In
a JT problem, the BO treatment starts with the diagonalization of
(\ref{real interaction Hamiltonian}) in the electronic degenerate space, at
fixed distortion field $\vec q$, i.e.\ the diagonalization of the matrix
$\Xi=\sum q_m V^{(m)}$.  Each electronic eigenvector $|\psi_\xi \rangle$ of
$\Xi$, of eigenvalue $\lambda_{\xi}$, generates a BO potential sheet
$V_{\xi}(\vec q)= \hbar\omega \left[\frac 12 \sum_m q_m^2 + \frac
g{\sqrt{2}} \lambda_{\xi}(\vec q) \right]$, including the harmonic
potential of the free vibrations.  At strong enough coupling $g$, the
separation of the potential sheets becomes large enough that the adiabatic
motion can be assumed to always follow the lowest BO potential sheet, while
virtual electronic excitations may be treated as a small correction.  For
Hamiltonians of type (\ref{ham sferica}), the set of points $\vec q_{\rm
min}$ of minimum potential energy, i.e.\ the classical stable
configurations, constitutes a continuous manifold, often denominated
Jahn-Teller manifold (JTM), and the value of the lowest BO potential there
is the classical JT stabilization energy $E_{\rm clas}=V_{\rm min}(\vec
q_{\rm min})$.

On the other side, due to time-reversal invariance of $H$, the space of all
possible (normalized) electronic eigenstates can be
represented by an (hyper-)sphere in the $[2L+1]$-dimensional real space (see
Fig.\ \ref{pathcl:fig}).  The BO dynamics realizes an adiabatic mapping of
the vibrational space into this electronic space.\cite{Ceulemans}  Indeed,
every point $\vec{q}$ on the JTM (in the vibrational space) is associated
to the electronic wave function $|\psi_{\rm min} (\vec{q})\rangle$,
corresponding to the lowest eigenvalue $\lambda_{\rm min}$ of the
electron-vibron interaction matrix $\Xi$:
\begin{equation}
\Xi(\vec{q})~ |\psi_{\rm min}(\vec{q})\rangle =
\lambda_{\rm min}(\vec{q})  |\psi_{\rm min}(\vec{q})\rangle \ .
\label{eigenvalue equation}
\end{equation}

This adiabatic mapping is two-valued, since opposite points $\pm |\psi_{\rm
min}(\vec{q})\rangle$ on the electronic sphere give the same JT
stabilization energy, thus corresponding to the same optimal distortion on
the JTM.  This identification of the antipodal points through the mapping
is the mechanism allowing the JTM to have different
connectedness\cite{Dubrovin} from the electronic sphere.  The latter is of
course simply connected, i.e.\ any closed path, or loop, on it can be
smoothly contracted to a single point.  The JTM, instead, may well be
multiply connected, i.e.\ it can have intrinsic ``holes'' in its topology.
In particular, in addition to the contractable loops (such as that mapping
on $\pi_1$ on the electronic sphere sketched in Fig.\ \ref{pathcl:fig}), on
the JTM we have the nontrivial class of those loops mapped on a path going
from a point to its antipode on the electronic sphere, such as $\pi_2$ in
Fig.\ \ref{pathcl:fig}.  In the traditional DJT systems ($E\otimes e$,
$T\otimes h$), these two classes of paths are topologically distinct: loops
belonging to class 2 may never be smoothly deformed into loops belonging to
class 1, whence the multiple-connectedness property of the JTM.  We see
therefore that this multiple connectedness is intimately related to the
mapping between the JTM and the electronic sphere.  The electronic sign
change characterising the loops belonging to class 2 is a case of Berry
phase\cite{Berry}.

When also the vibrational motion is quantized, the overall (vibronic) BO
wave function is factorized in the direct product of the electronic
adiabatic state times the wave function for the slow degrees of freedom
$\vec q$: since the vibronic wave function is a regular, single-valued
function, the $\vec q$ degrees of freedom must cope with the electronic
phase change, which acts as a special boundary condition to quantization.
As a consequence the motion on the JTM is constrained by special selection
rules.  For example the JTM of the simple $E\otimes e$ system is a circle:
the low-energy vibronic spectrum is indeed a $j^2$ spectrum as for a
circular rotor, but the Berry phase implies $j=\pm\frac 12,\pm\frac
32,...$, instead of $j=0,\pm 1,\pm 2,...$ as for an ordinary quantum
rotor.\cite{Bersuker,Koizumi}  Similarly, the JTM of the $T\otimes h$
(i.e.\ ${\cal D}^{(1)} \otimes d^{(2)}$, in the spherical language) is
equivalent to a sphere,\cite{AMT,ob71} but out of all the states, labeled
by $J,M$, of a particle on a sphere, the Berry phase keeps only the odd-$J$
ones.\cite{Bersuker,AMT,ob71}  Note in particular that in these examples
the presence of a Berry phase rules out a nondegenerate ground state.  The
same symmetry of the degenerate electronic state, prior to the vibronic
coupling/distortion, is enforced to the strong coupling DJT ground state.

\section{Absence of the Berry Phase}
\label{nobp:sect}

As anticipated above, though a very frequent phenomenon, the Berry phase is
{\em not} automatically implied by linear JT Hamiltonians (\ref{2nd quant
ham}).  The above discussion should make clear that the absence of the
Berry phase in a DJT system is linked to a mechanism leading to equivalence
of the paths in the class 1 and 2.  This mechanism should also coexist with
the two-valued adiabatic mapping sketched in the previous section.
The solution of the
riddle is provided by a point $\vec q_d$ {\em on the JTM} where the mapping
is degenenerate, i.e.\ it links $\vec q_d$ not just to a pair of opposite
points $\pm |\psi_{\rm min}(\vec q_d)\rangle$ on the electronic sphere, but
to the whole circle of linear combinations $\cos \theta ~|\psi_1(\vec
q_d)\rangle+ \sin \theta ~|\psi_2(\vec q_d)\rangle $ of two degenerate
electronic eigenstates (such as, for example, $\Delta$ in
Fig.\ \ref{pathcl:fig}).  If such a point is present, any loop of class 2 on
the JTM may be deformed smoothly so that its image on the electronic sphere
becomes half this circle, thus the single point $\vec q_d$ on the JTM:
class 2 loops are therefore equivalent to class 1 (contractable) loops,
and, therefore, the JTM is simply connected.  No Berry phase is possible in
such a case.\cite{Paris97}

The possibility of such degenerate points may be considered {\it via}
careful analysis of the structure of the multiple BO potential sheets
$V_{\xi}$ given by the eigenvalues $\lambda_{\xi}$ of the
electron-vibration interaction operator.  In most ``classical'' cases of
Berry-phase--entangled linear JT systems ($E\otimes e$, $T\otimes h$, ...),
the lowest potential sheet remains separated from the next lowest
one by a finite gap throughout the JTM.  {\em Conical} intersections
between these two sheets take place at some point in the distortion space,
far from the potential minimum.  In the $E\otimes e$, for example, such a
point is the origin $\vec q =0$.\cite{nodegsheet:note}  There is, however,
a second possiblilty: somewhere on the JTM, the lowest BO potential sheet
gets {\em tangentially} degenerate to the next lowest BO potential
sheet.  The energy difference between the two lowest levels is quadratic
in the distance from the degenerate point.

This possibility is indeed realized in our spherical model.  Take for
example the case $L=2$, $l=2$ and consider the point $\vec q_d=\left(0,0,-
\frac 1{\sqrt{7}} g,0,0 \right)$.\cite{Delos96}  Simple inspection of the
coupling matrix $\Xi$ (\ref{V0}) shows that its lowest eigenvalue $-|q_d|
\gamma^{+}_{2,-2} \equiv -|q_d| \langle l\;0 | L\;2, L\;-2 \rangle=
-\frac{\sqrt{2}}{7}g$ is twofold degenerate.  The classical BO potential
energy is $E_{\rm clas}=V_{\rm min}(\vec q_d) =-\frac 1{14} g^2 \hbar
\omega$, and no other $\vec q$ yields lower potential energy than this: by
definition, $\vec q_d$ does belong to the JTM.  By inspection of Eqs.\
(\ref{V0}-\ref{V2}), it is possible to verify that the direct coupling
between the two degenerate electronic states (element $\Xi_{\pm 2,\mp 2}$
of the coupling matrix) vanishes identically: moving away from $\vec q_d$,
these two states remain degenerate to first order, and the degeneracy is
only lifted by indirect, second order coupling to the other states.  $\vec
q_d$ is indeed a tangency point.  With this, we have detailed the mechanism
for the absence of a Berry phase in the ${\cal D}^{(2)}\otimes d^{(2)}$
model.

It is natural to search possible extensions of this mechanism to other
cases of spherical DJT models (\ref{ham sferica}).  Eqs.\
(\ref{V0}-\ref{V4}) show that on the $\hat q_0$ axis, the electronic states
are pairwise degenerate (except for the $\tilde{c}^\dagger_0 |0\rangle$
state).  We are interested in the pair corresponding to the lowest and next
lowest BO sheet, thus to the maximum numerical value of the Clebsch-Gordan
$|\gamma^+_{m,-m}|$.  For the ${\cal D}^{(2)}\otimes d^{(2)}$ described
above, the maximum is indeed $\gamma^+_{2,-2}$.\cite{other_eigenvalue:note}
On the contrary, for the ${\cal D}^{(2)}\otimes d^{(4)}$ case, the largest
coefficient is $\gamma^+_{0,0} = \left(\frac {18}{35}\right)^{1/2}$,
corresponding to the singlet state.  Thus on the $\hat q_0$ axis there are
no degeneracy points of the two lowest BO sheets.  We verified that this
case has no degenerate points anywhere on the JTM, the gap between the
lowest and next lowest BO sheet being a constant $\frac g{\sqrt{2}} |q_d|
\left(\frac 5{14}\right)^{1/2}=\frac 3{14} g^2$ across all the JTM.  We
conclude that the ${\cal D}^{(2)}\otimes d^{(4)}$ is not a
Berry-phase--free model.  The same applies to all ``complete'' systems,
i.e.\ the ${\cal D}^{(L)}\otimes d^{(l=2L)}$, since $\langle 2 L\;0 | L\;0,
L\;0 \rangle$ (corresponding to the nondegenerate electronic state) is the
largest Clebsch-Gordan coefficient among all $\langle 2 L\;0 | L\;M, L\;-M
\rangle = N(L)/(L-M)!(L+M)!$ (where $N(L)$ is a function of $L$ only).

If we consider now the ${\cal D}^{(L)}\otimes d^{(l=L)}$ models, we see
that the coefficients $|\langle L\;0 | L\;M, L\;-M \rangle|$ are peaked
around $|M|=\hat M\approx 86\% L$ (for large $L$).  For $L<50$, say, the
peak is rather sharp, allowing to concentrate on the $2\times 2$ block of
the interaction matrix:
\begin{equation}
\small
\left(\begin{array}{cc}
\Xi_{-\hat M,-\hat M}	& \Xi_{-\hat M,\hat M} \\
\Xi_{\hat M,-\hat M}	& \Xi_{\hat M,\hat M}
\end{array} \right)
= \left(\begin{array}{cc}
\chi_{\hat M,-\hat M} q_0	& 0 \\
0 &	\chi_{-\hat M,\hat M} q_0
\end{array} \right)
\label{2x2 con coord}
\end{equation}
Other diagonal and off-diagonal contributions in the block vanish, since
they could only be related to $q_{\pm 2\hat M}$, which does not exist,
since $2\hat M>l=L$.  This block, therefore, gives a twofold degenerate
electronic ground state, with second-order departures from the degeneracy,
as the distortion moves away from the $\hat q_0$ axis.  We conclude that
the two lowest BO sheets, in all the ${\cal D}^{(L)}\otimes d^{(L)}$
models, get tangent at (at least) one point, thus making equivalent all
loops on the JTM.  As a consequence, this class of models must be
considered as Berry-phase--free.

Finally, we apply the same reasoning to the study of a generic ${\cal
D}^{(L)}\otimes d^{(l)}$ model.  For $l<L$ the $|M|=\hat M$, at which the
relevant coefficients $|\langle l\;0 | L\;M, L\;-M \rangle|$ is maximum,
always satisfies the inequality $2\hat M>l$.  Thus, these are yet more
Berry-phase--free systems.  On the contrary, for $l>L$, we go from a case
without (${\cal D}^{(L)}\otimes d^{(L)}$) to a case with (${\cal
D}^{(L)}\otimes d^{(2 L)}$) Berry phase, thus the result is not
trivial. For large enough $l$, the maximum coefficient is attained at some
$\hat M\leq l$, and this leaves nonzero direct off-diagonal coupling
elements in the $2\times 2$ block (\ref{2x2 con coord}).  In such a case,
the degenerate point is not a tangency, but a conical intersection, thus it
cannot belong to the JTM.  It can be verified numerically that the tangency
point on the $\hat q_0$ axis disappears for all $l\geq l_c(L)$, where the
``critical value'' $l_c(L)$ is found between $L$ and $2L$.  For
$L=2,3,4,... 10$, $l_c(L)$ takes values $4, 4, 6, 8, 8, 10, 12, 12, 14$
respectively: thus, for example, a ${\cal D}^{(7)}\otimes d^{(8)}$ model
has no Berry phase, while ${\cal D}^{(9)}\otimes d^{(14)}$ can have it.
Our demonstration focuses on the $\hat q_0$ axis: we cannot rule out the
possibility of other tangency points elsewhere on the JTM.  Thus the
presence of the Berry phase is verified only for $l=2 L$, where the gap
between the two lowest sheets is a constant on the JTM, while it is only
likely for $l_c(L)\leq l<2L$.

\section{Numerical results}
\label{numerical:sect}

We have classified Berry-phase--wise a large class of spherical DJT
systems: it should be possible to evidence the signatures of this property
in their vibronic spectrum.  It was previously shown\cite{Delos96} that the
ground state of the ${\cal D}^{(2)}\otimes d^{(2)}$ system becomes
nondegenerate at strong coupling.  We have applied the same technique of
numerical diagonalization on a truncated basis to the next
Berry-phase--free model, ${\cal D}^{(4)}\otimes d^{(4)}$.  In Fig.\
\ref{spec4x4:fig}, we show the vibronic energy of the lowest $L=0$ and
$L=4$ vibronic states above the BO potential minimum $E_{\rm clas}$: the
lowest nondegenerate and degenerate states cross at $g\approx 8$,
the former becoming the strong-coupling ground state, as the absence of a
Berry phase predicts.

However, an analogous test for the ${\cal D}^{(4)}\otimes d^{(2)}$ model
finds a degenerate $L=4$ ground state up to coupling $g=20$.  An
explanation could be searched in the nature of the lowest $L=0$ state at
weak-coupling: while in ${\cal D}^{(L)}\otimes d^{(L)}$ models it comes as
a fragment of the one-vibron multiplet, in this case it originates from the
two-vibron multiplet, due to angular momentum conservation.  It seems as if
the angular moment of the $l=2$ vibration was insufficient to screen the
large electronic moment, thus, even in absence of Berry phase, the ground
state remains degenerate.

\section{Discussion}
\label{discussion:sect}

In the Berry-phase-free DJT systems, the tangency of the lowest two BO
sheets, strictly speaking, invalidates the BO treatment, which assumes a
large gap between the lowest electronic state and the next one.  Thus,
paradoxically, in these systems it is not the Berry phase which is related
to a breakdown of the BO approximation, but its {\em absence}.  Indeed, we
have shown that, even though these degeneracies are present only locally on
the JTM, they affect radically the whole coupled dynamics.

Our analysis considers spherical DJT models: however, it can be extended to
molecular point groups, by substituting the relevant Clebsch-Gordan
coefficients.\cite{butler81} There are cases, such as the linear $T\otimes
h$ in icosahedral symmetry (or equally coupled $T\otimes (e+t)$ in cubic
symmetry) which are equivalent to some spherical models, but of course this
is not always the case, and other surprising results could be found by
looking for tangencies in different systems.

Also, we assume a linear JT coupling scheme (Hamiltonian (\ref{2nd
quant ham})), which is the less realistic, the stronger the JT distortion.
The perturbative introduction of higher-order couplings has usually effects
similar to those produced by different linear coupling $g_E\neq g_T$ in
cubic symmetry,\cite{ob69} i.e.\ of ``warping'' the JT trough.  The
continuous JTM becomes a set of isolated minima, connected by low-energy
paths passing through saddle points.  The symmetry of the Hamiltonian and
of the JTM is reduced to the symmetry group $G$ of the molecule.  Yet, the
connectedness properties are topological properties, therefore they are
robust against perturbations such as the potential warping: even if the
symmetry is reduced, the Berry phase is still present or absent as
determined by the linear part.  Ham\cite{Ham87} has shown for the $E
\otimes e$ coupling scheme that the introduction of quadratic terms in the
Hamiltonian does not substantially change the picture as far as the Berry
phase and the degeneracy of the ground state are concerned.  In fact, even
at strong JT coupling, the tunneling among rather deep isolated minima is
affected by the electronic phase,\cite{Bersuker} and, as a result, the
lowest tunnel-split state retains the same symmetry and degeneracy as in
the purely linear-coupling case.  Of course, in the extreme limit of very
large distortion, higher-order terms dominate, and the DJT is replaced
by a static distortion, where any Berry phase argument becomes irrelevant.

\section{Conclusion}
\label{conclusion:sect}

In summary, we propose evidence for a whole family, following the ${\cal
D}^{(2)}\otimes d^{(2)}$, of Berry-phase--free dynamical JT systems: these
are the ${\cal D}^{(L)}\otimes d^{(l)}$ models, with $l<l_c(L)$, where the
critical value $l_c(L)$ lies between $L$ and $2L$.  For these models, we
also show that the absence of the Berry phase does not automatically imply
a nondegenerate strong-coupling vibronic ground state.  Moreover, we prove
that a Berry phase is present in the ${\cal D}^{(L)}\otimes d^{(2L)}$
models, which, as a consequence, have a ${\cal D}^{(L)}$ $[2L+1]$-fold
degenerate ground state for any coupling.

\section*{Acknowledgement}

We thank M.\ Altarelli, E.\ Tosatti and Lu Yu for useful discussions.

\begin{figure}
\epsfxsize 10.0cm
\inseps{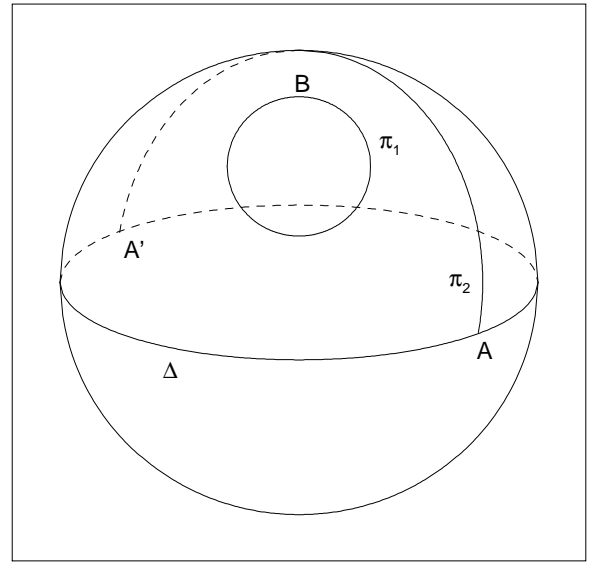}{0.6}
\caption{ A sketch of the electronic sphere.  The picture
individuates the two classes of paths mapping onto closed loops in the JTM:
paths of the type $\pi_2$ involve a sign change (from A to A') of the
electronic state (a Berry phase).  These two types of paths can be
distinguished in all $N\geq 3$-dimensional cases.
\label{pathcl:fig}}
\end{figure}\noindent

\begin{figure}
\epsfxsize 10.0cm
\inseps{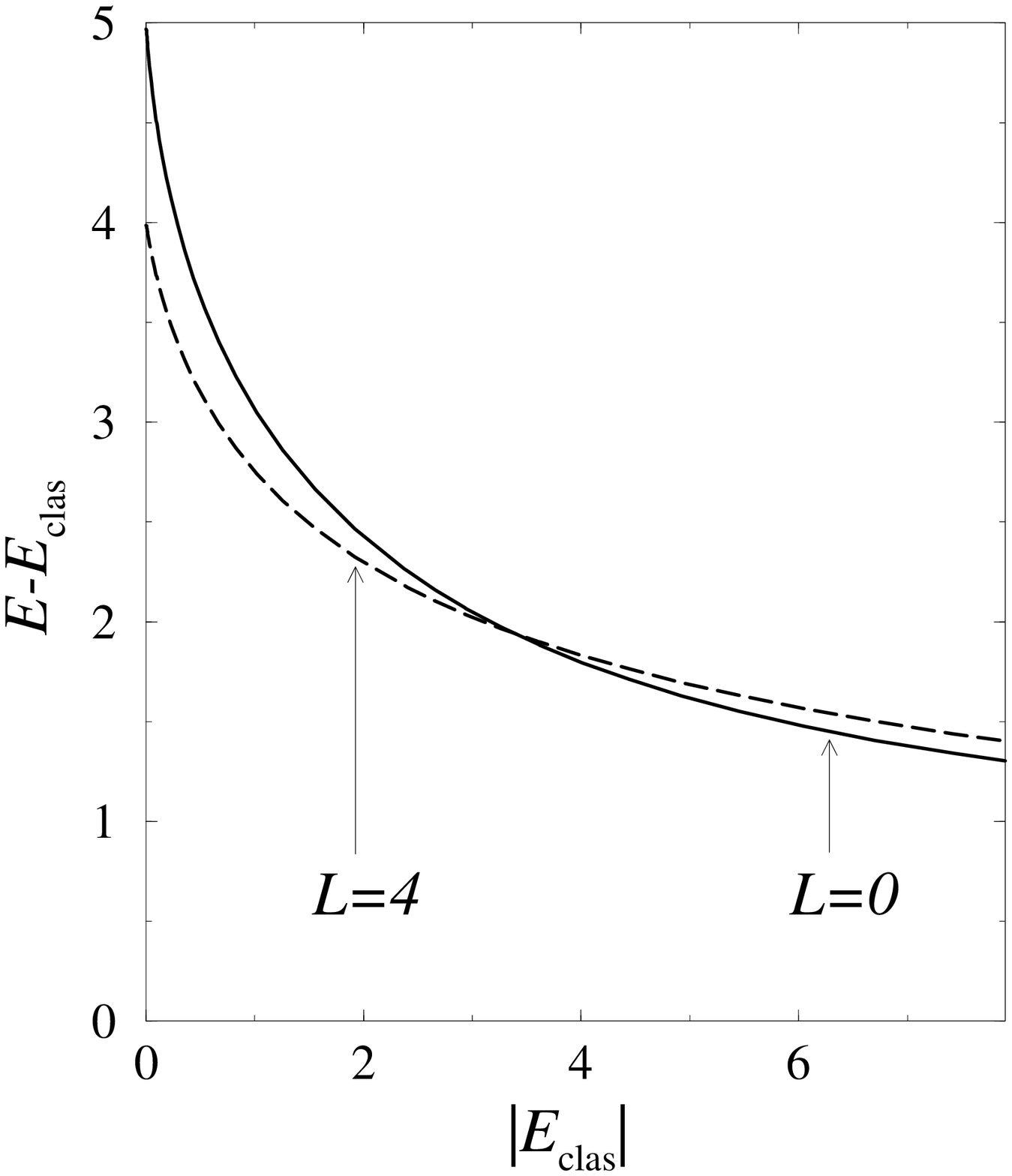}{0.5}
\caption{ The energy of the lowest $L=0$ and $L=4$ vibronic states of the
${\cal D}^{(4)}\otimes d^{(4)}$ system, as a function of $|E_{\rm
clas}|=\frac {63}{1144} g^2 \hbar\omega$.  The residual zero-point energy
of $\frac 12 \hbar\omega$ is subtracted.  The energies, in units of the
harmonic quantum $\hbar\omega$, are obtained by exact diagonalization of
the Hamiltonian (\protect\ref{ham sferica}) on a truncated Hilbert space
including up to 16 boson states, enough to reach convergence in this range
of couplings.
\label{spec4x4:fig}}
\end{figure}\noindent

\end{document}